\documentclass[aps,prb,twocolumn,amsmath,showpacs,amssymb,superscriptaddress]{revtex4}
\usepackage{amsmath}
\usepackage{graphicx}
\usepackage{amssymb}
\usepackage{dsfont}
\usepackage{color}

\newcommand{\pdag}{{\phantom{\dagger}}}

\begin{document}

\title{Time reversal symmetry broken fractional topological phases at zero magnetic field}

\author{Tobias Meng}
\affiliation{Department of Physics, University of Basel, Klingelbergstrasse 82, CH-4056 Basel, Switzerland}
\author{Eran Sela}
\affiliation{Raymond and Beverly Sackler School of Physics and Astronomy, Tel-Aviv University, Tel Aviv, 69978, Israel}

\begin{abstract}
We extend the coupled-wire construction of quantum Hall phases, and search for fractional topological insulating states in models of weakly coupled wires at zero external magnetic field. Focussing on systems beyond double copies of fractional quantum Hall states at opposite fields, we find that spin-spin interactions can stabilize a large family of fractional topological phases with broken time reversal invariance. The latter is manifested by spontaneous spin polarization, by a finite Hall conductivity, or by both. This suggests the possibility that fractional topological insulators may be unstable to spontaneous symmetry breaking. 
\end{abstract}

\maketitle

\section{Introduction}
Symmetry protected topological phases  have sprung to the forefront of condensed matter physics. The impetus for such an explosion of interest began with the theoretical prediction\cite{kane_mele_05,bernevig_zhang_2006,bernevig_hughes_zhang_2006,liu_08} and observation of two-dimensional topological insulators\cite{ti_review} in HgTe/CdTe\cite{koenig_07} and InAs/GaSb\cite{knez_11,du_13} heterostructures.
From there, the field has now spread to encompass interaction induced topological phases, including in particular fractional topological insulators.~\cite{levin_stern_09}

Topological insulators can be understood as a time-reversal symmetric generalization of a quantum Hall state to a bilayer system in which the two layers, physically corresponding to spin up and spin down electrons, act as if they were subject to opposite magnetic fields $B_{SO} \hat{z} s_z$. This kind of physics can originate from spin-orbit coupling.\cite{kane_mele_05,bernevig_zhang_2006,bernevig_hughes_zhang_2006,liu_08} A topological insulator has helical edge modes (two modes related by time reversal symmetry), instead of the chiral edge mode of a quantum Hall state (a single mode whose direction of motion is dictated by the time reversal symmetry breaking magnetic field).

In the presence of electron-electron interactions, fractionalized versions of topological insulating states have been predicted.\cite{levin_stern_09}
A simple way to generalize the above construction is to imagine that the two spin species each form a fractional quantum Hall (FQH) state. This situation can be realized in a
toy model similar to the one described above. The only
new element is a short-range two-body interaction between electrons of the same spin, while electrons of
different spin do not interact at all. This toy model can then
be mapped to two decoupled FQH systems of the same
filling factor, and  subject to opposite magnetic fields.\cite{Freedman,levin_stern_09}

More exotic fractional topological insulating phases may occur due to interactions between the two spin species. So far, a number of time-reversal symmetric phases arising due to complex inter spin interactions have been predicted.\cite{bernevig_zhang_2006,levin_stern_09,neupert_11,santos_11,levin_stern_12} However, it should be kept in mind that, much like the Stoner instability of a metal to a spin polarized state, which is driven by exchange interactions, strong interactions between the two spin species in a topological insulator may result in a time-reversal  symmetry broken phase.

To address such phases, we study a model of weakly coupled interacting spinful wires. Similar constructions based on arrays of one-dimensional subsystems have proven a powerful approach for the description of integer, fractional, and more exotic quantum Hall states,\cite{poilblanc_87,yakovenko_91,sondhi_00,kane_02,teo_2014,klinovaja_epjb_14,GANGOF11,OregSelaStern} for the analysis of fractional topological insulators effectively consisting of two decoupled quantum Hall layers,\cite{sagi_14,klinovaja_14} as well as for an alternative general classification of topological states.\cite{neupert_14} Experimentally, arrays of coupled quantum wires could for instance be engineered in epitaxially grown multilayer systems using the cleaved edge overgrowth method.\cite{auslaender_02} In addition, quantum Hall physics have been identified in Bechgaard salts,\cite{poilblanc_87,yakovenko_91} which consist of effectively two-dimensional arrays of coupled one-dimensional subsystems.

Different from the coupled wire construction of a fractional quantum Hall state, there is no external magnetic field in the main part of our analysis. Instead, we consider a model with Rashba type spin-orbit coupling which increases linearly from wire to wire, mimicking the spin-dependent magnetic field $B_{SO} \hat{z} s_z$.  Without interactions between the two spin species, this model trivially reproduces the pair of decoupled FQH systems at opposite magnetic fields.\cite{Freedman,levin_stern_09}

As a warmup for the construction of topological states with strong interactions between the two spin species, however, we first reconsider double layer quantum Hall states at finite magnetic field $B \hat{z}$. These are described by generalizations of Laughlin's wave functions suggested by Halperin,\cite{wen_book,halperin_83} and intimately related to the Haldane-Halperin hierarchy states.\cite{haldane_83,halperin_84} The wave functions are characterized by integers $(m_{\uparrow}m_{\downarrow}n)$, with $m_\uparrow$ and $m_\downarrow$ odd, and read
\begin{align}
&\Psi_{m_{\uparrow}m_{\downarrow}n}(z_1,\ldots, z_{N_{0\uparrow}},w_1,\ldots, w_{N_{0\downarrow}}) = \label{eq:halp_wave}\\
&\prod_{i<j}(z_i-z_j)^{m_{\uparrow}}\,\prod_{p<q}(w_p-w_q)^{m_{\downarrow}}\,\prod_{r,s}(z_r-w_s)^{n}\nonumber\\
&\times e^{-\sum_i (|z_i|^2+|w_i|^2)/4 l_b^2}~.\nonumber
\end{align}
Here, the magnetic length is $l_b = \sqrt{\hbar/(e B)}$, the two layers are labeled by the (pseudo-) spin $\sigma = \uparrow,\downarrow$, and  the complex numbers $z_k = x_{k\uparrow}+iy_{k\uparrow}$ and $w_k=x_{k\downarrow}+iy_{k\downarrow}$ are defined by the $x$ and $y$ coordinates of the electrons in the two layers. The latter have fillings $\nu_{\sigma}$ discussed in Eq.~\eqref{eq:fillings_halperin} below, and contain $N_{0\sigma}$ electrons. The factor $\prod(z_r-w_s)^{n}=\prod([x_{r\uparrow}-x_{s\downarrow}]+i[y_{r\uparrow}-y_{s\downarrow}])^{n}$ encodes the inter layer correlations. The wire construction for these Halperin states is formulated in Sec.~\ref{sec:halpering_states}, reproducing all of their topological properties including filling factors, quasiparticle charges and edge structure.

We then move in Sec.~\ref{sec:gen_halperin} to the main problem of interest at zero total magnetic field $B=0$, but with spin-orbit coupling, in which case the two layers, now corresponding to the two spin species, have effectively opposite magnetic fields $B_{SO }s_z \ne 0$. Repeating the wire construction in this case leads to states that show different properties than the Halperin states. They can, however, still be labeled by three integers $(m_\uparrow m_\downarrow n)$ with odd $m_\uparrow$ and $m_\downarrow$. These integers relate to the filling fractions as
\begin{align}
\label{fillingfactor}
\nu_\uparrow =\frac{\rho_\uparrow h}{e B_{SO}} =  \frac{m_\downarrow-n}{m_{\uparrow}m_\downarrow+n^2}, \nonumber \\
 \nu_\downarrow =\frac{\rho_\downarrow h}{e B_{SO}} = \frac{m_\uparrow+n}{m_{\uparrow}m_\downarrow+n^2}~.
\end{align}
 Note that in our notation both $\nu_\uparrow$ and $\nu_\downarrow$ are positive, as they are proportional to the respective electron densities $\rho_\uparrow$ and $\rho_\downarrow$. The integers $(m_\uparrow m_\downarrow n)$ should be chosen accordingly. An important implication of Eq.~(\ref{fillingfactor}) is  that when the two spins (or layers) are sufficiently strongly coupled to result in $n\neq0$, the two filling factors can be different from each other, $\nu_\uparrow  \ne \nu_\downarrow$, which translates to a finite magnetization. After deriving these states, we discuss their physical properties, including the edge structure, bulk quasiparticle excitations, and the associated Hall conductivity. The latter vanishes by symmetry in a time reversal invariant system. The general states considered in this work, however, have $\sigma_{xy}= \frac{e^2}{h}  \frac{2n + m_\downarrow-m_\uparrow}{m_\uparrow m_\downarrow + n^2}$. For a finite $n$, we thus find that the system is either spin polarized, or has a finite Hall conductivity, or both. This indicates that time reversal symmetry is spontaneously broken in these states.

Similar to the time reversal symmetric case, the edge structure of these states consists of two counter propagating modes. If one assumes spin conservation (conservation  of $s_z$), these counter propagating edge modes are conserved.  In the case of broken time reversal symmetry, and more generally when $s_z$ is not conserved due to spin-orbit coupling, it is, however, possible to gap out the counter propagating edge modes by perturbations such as impurities.

Consider as an example states with $m_\uparrow = m_\downarrow \equiv m$. For $n=0$, the two spins are decoupled. In this case, $\nu_\uparrow = \nu_\downarrow=1/m$ correspond to Laughlin states for both spins, but at opposite magnetic fields. For $n \ne 0$ the filling factors are different from each other, and do not correspond to decoupled FQH states for the two spins. In a (331) state, for example, the total filling of $\nu =\nu_\uparrow  +\nu_\downarrow  =\frac{3}{5}$ decomposes spontaneously as $\frac{3}{5}=\frac{1}{5}+\frac{2}{5}$ between the spins. The choice of the sign of $n$, which does not change the total filling factor, represents the spontaneous symmetry breaking in this state. Besides the finite magnetization, the symmetry breaking also leads to a non-zero Hall conductivity $\sigma_{xy} = \frac{e^2}{h}  \frac{2n}{m^2+n^2}$. As for the magnetization, the sign of $\sigma_{xy}$ is given by the sign of $n$.

As a second example, consider states with $m_\uparrow = m-n$ and $m_\downarrow = m+n$, where the parity of $m$ must be opposite to that of $n$ to guarantee that $m_\sigma$ are odd. For any $n$, the individual filling factors are given by $\nu_\sigma = 1/m$. Let us focus on $\nu_\uparrow=\nu_\downarrow = 1/3$ for concreteness. For these filling factors, the system could for instance be in a $(330)$ state or a $(152)$  state (the latter is related by time reversal to the $(51-2)$ state). Which of these states is the most stable one depends on the microscopic interactions in the system. While now, the symmetry breaking is not reflected in a finite spin polarization $\nu_\uparrow-\nu_\downarrow$ anymore, it still gives rise to a Hall conductivity $\sigma_{xy} = \frac{e^2}{h}  \frac{4n}{m^2}$, which equals $\pm \frac{e^2}{h} \frac{8}{9}$ for the $(152)$ or $(51-2)$ states, respectively.

Notice that one can also find states with vanishing Hall conductivity, but with a finite spin polarization. Those are the $(m+n, m-n,n)$ states (with $n \ne 0$). Thus, in the general time reversal symmetry broken phases considered here, the Hall conductivity or the spin polarization can vanish - but not both of them.

In Sec.~\ref{sec:conclusion}, we finally conclude with a comparison to other approaches for fractional topological insulators.\cite{levin_stern_09,neupert_11,santos_11,levin_stern_12} Our results are consistent with works based on $K-$matrix Chern-Simons theories,  which find that only substantially more delicate states may still preserve time reversal symmetry and at the same time have strong correlations between the two spin species. The significance of our analysis is thus that time reversal broken states are natural competing phases which should be considered in the search for spin-correlated fractional topological insulators.

\section{Coupled wire construction of an $(m_{\uparrow}m_{\downarrow}n)$-type Halperin state in a bilayer system}\label{sec:halpering_states}
In this section, we provide a wire construction for double layer systems. We obtain an alternative formulation for the well known $(m_{\uparrow}m_{\downarrow}n)$ Halperin states given by the wave function in Eq.~(\ref{eq:halp_wave}), and describing two quantum Hall liquids on two layers with filling factors
\begin{align}
\begin{pmatrix}
\nu_{\uparrow}\\ \nu_\downarrow
\end{pmatrix} =
\frac{1}{m_\uparrow m_\downarrow - n^2}\begin{pmatrix}m_{\downarrow}-n
\\m_{\uparrow}-n
\end{pmatrix}~.\label{eq:fillings_halperin}
\end{align}
Note that stability of this state against phase separation requires $m_{\uparrow}m_{\downarrow}-n^2\geq0$ (the case $m_\uparrow m_\downarrow =n^2$ corresponds to a fully symmetric spin wave function with ill-defined individual spin occupations).\cite{degail_08} Eq.~\eqref{eq:fillings_halperin} encodes that the filling decreases when the inter layer correlations are enhanced: an inter layer repulsion pushes all electrons further apart.

\subsection{Wire construction}\label{subsec:wire_const}
\begin{figure}
  \centering
  \includegraphics[width=\columnwidth]{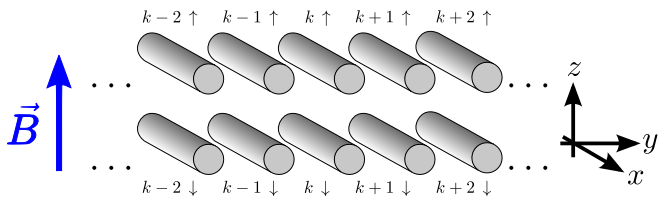}
  \caption{The considered setup: two stacked arrays of quantum wires containing spinless electrons. The layers are labeled by an index $k$ within each plane, and a (pseudo-) spin $\sigma = \uparrow,\downarrow$ distinguishing the two  planes. The entire system is subject to a homogenous magnetic field $\vec{B}$ perpendicular to the plane of the arrays.}
  \label{fig:array}
\end{figure}

\begin{figure}
  \centering
  \includegraphics[width=\columnwidth]{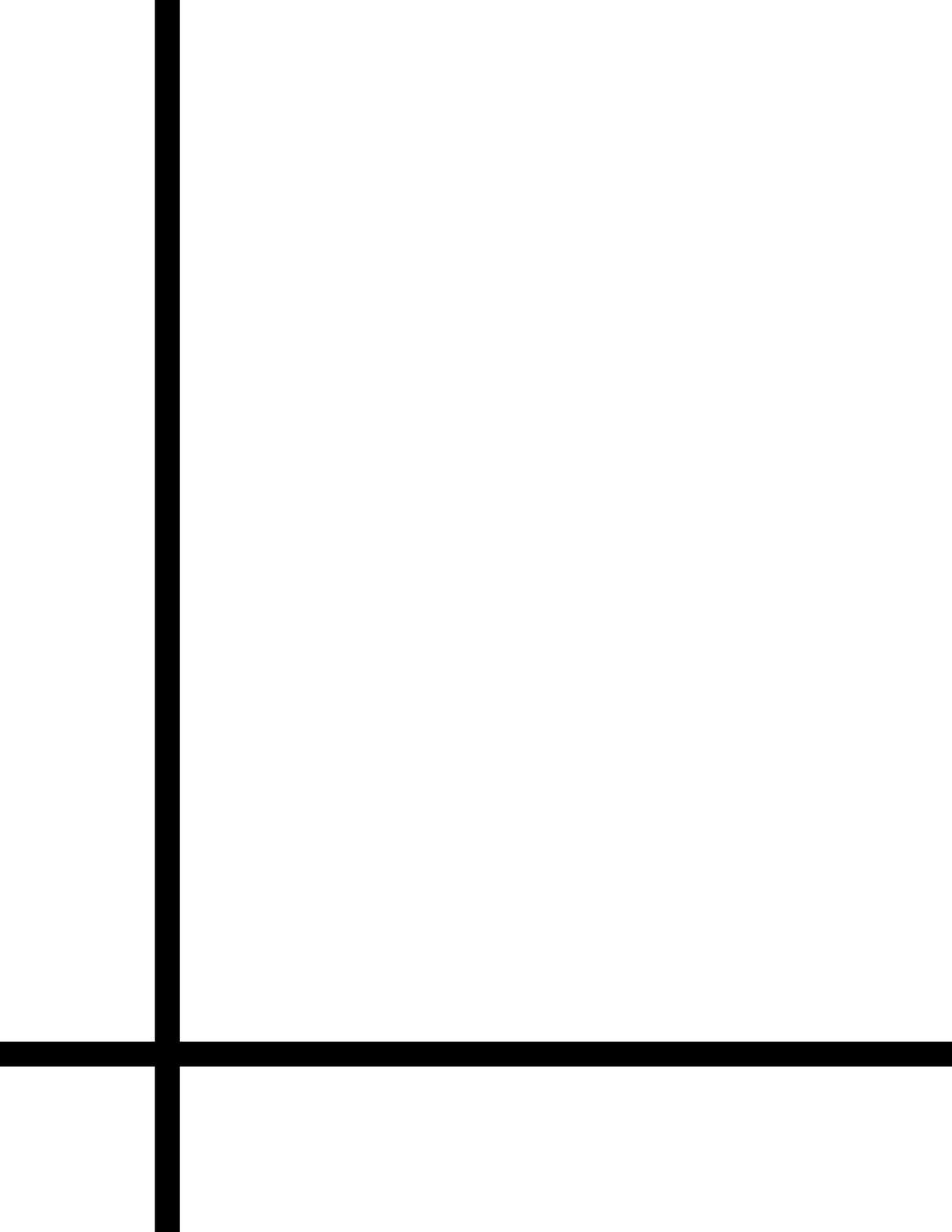}
  \caption{The dispersions $E(p)$ as a function of the momentum $p$ in the $\hat{x}$-direction of the bilayer system (in a representation based on the Landau gauge). The wires are labeled as in Fig.~\ref{fig:array}. The chemical potential in layer $\sigma=\uparrow,\downarrow$, denoted by $\mu_\sigma$, translates into a filling factor $\nu_\sigma$. As discussed in the main text, the momentum shift between neighboring wires is $\Delta p = e B a=2p_{F\sigma}/\nu_\sigma$.}
  \label{fig:dispersions}
\end{figure}

For the wire construction of Halperin states, we consider the setup depicted in Fig.~\ref{fig:array}, namely a large array of quantum wires arranged in two layers. Each layer consists of $N$ wires. The latter contain spinless electrons, and are subject to a magnetic field $\vec{B} = B\,\hat{z}$ perpendicular to the plane of the layers. We label the wires by $k=1,\ldots,N$, the wire number within each layer, and $\sigma=\uparrow,\downarrow$, the pseudospin distinguishing the two layers. Since we consider spinless electrons, the magnetic field affects the system by its orbital effect. It is convenient to work in the Landau gauge, in which $\vec{B} = \vec{\nabla}\times\vec{A}$ with $\vec{A}=(-By,0,0)^T$. In this gauge, the momenta along the $\hat{x}$-direction, denoted by $p$, are effectively shifted by $\Delta p = e B a$ for neighboring wires, where $a$ is the distance between the wires (taken to be identical in both layers). This leads to the dispersion relation shown in Fig.~\ref{fig:dispersions}. Using the fact that the two-dimensional density of each layer satisfies $a\,n_{\sigma} = p_{F\sigma}/(\pi \hbar)$, and that $\nu_\sigma \equiv n_\sigma h/(eB)$, the momentum shift can be expressed as $\Delta p = 2 p_{F\sigma}/\nu_{\sigma}$, where $p_{F\sigma}$ is the Fermi momentum of a given wire in layer $\sigma$ measured with respect to the minimum of its dispersion. The difference between the right and left Fermi point in neighboring wires is thus $2 p_{F\sigma}/\nu_{\sigma}-2p_{F\sigma} = (1-\nu_{\sigma})\,\Delta p$, as indicated in Fig.~\ref{fig:dispersions}.

The electrons are annihilated by operators $\psi_{k\sigma}(x)$, which satisfy the usual anticommutation relation $\{\psi_{k\sigma}(x),\psi_{k'\sigma'}(x')\} = \delta_{kk'}\delta_{\sigma\sigma'}\delta(x-x')$. In the remainder, however, we will not use these fermionic operators, but treat the array of wires in the Luttinger liquid formalism. To this end, we first restrict the theory to low energy excitations close to the Fermi points, which gives rise to right ($R$) and left ($L$) moving modes.\cite{giamarchi_book} Measuring the momentum in each wire with respect to the minimum of its dispersion, and working from now on in units of $\hbar=1$, the right and left movers relate to the initial electronic operators as $\psi_{k\sigma}(x)\approx e^{-ip_{F\sigma}x}L_{k\sigma}(x)+e^{ip_{F\sigma}x}R_{k\sigma}(x)$. These are bosonized as $r_{k\sigma}(x) =(U_{r k\sigma}/\sqrt{2\pi\alpha})\,e^{-i\Phi_{r k\sigma}(x)}$, where $r=R,L \equiv +1,-1$, while $\alpha^{-1}$ is a large momentum cutoff, and with $U_{r k \sigma}$ being a Klein factor (which we drop as usual in the remainder since they are not important for our discussion). The chiral bosonic fields satisfy the commutation relation $[\Phi_{r k \sigma}(x),\Phi_{r' k' \sigma'}(x')]=\delta_{rr'}\delta_{kk'}\delta_{\sigma\sigma'}\,i\pi r\,\text{sgn}(x'-x)$. It is helpful to also define the fields $\phi_{k \sigma}(x) = (\Phi_{R k\sigma}(x)-\Phi_{L k\sigma}(x))/2$ and  $\theta_{k \sigma}(x) = (-\Phi_{R k\sigma}(x)-\Phi_{L k\sigma}(x))/2$, which have the commutation relation $[\phi_{k\sigma}(x),\theta_{k\sigma}(x')] = \delta_{kk'}\delta_{\sigma\sigma'}\,(i\pi/2)\,\text{sgn}(x'-x)$. The field $\phi_{k\sigma}$ relates to the integrated density of electrons in wire $k$ of layer $\sigma$, while $\theta_{k\sigma}$ is proportional to their integrated current. These definitions allow to express the Hamiltonian of the decoupled wires in a bosonized language as
\begin{align}
H_0=\sum_{k,\sigma}\int \frac{dx}{2\pi}\left[\frac{u_{k\sigma}}{K_{k\sigma}}(\partial_x\phi_{k\sigma})^2+u_{k\sigma}K_{k\sigma}(\partial_x\theta_{k\sigma})^2\right]~,\label{eq:arrayll}
\end{align}
where $u_{k\sigma}$ is the effective velocity in wire $k\sigma$, while $K_{k\sigma}$ is its Luttinger liquid parameter. The use of $u_{k\sigma}$ and $K_{k\sigma}$ allows one to take electron-electron interactions of density-density type into account.\cite{giamarchi_book} In Eq.~\eqref{eq:arrayll}, we have neglected density-density interactions between different wires, which could be included by a straightforward generalization of $H_0$.

\subsection{Allowed couplings and the Halperin states}
\label{se:couplingHalperin}
In the following, we analyze the kinds of inter wire couplings that stabilize a Halperin state. For hierarchical fractional quantum Hall states related to the Halperin states of $(mmn)$ type, a coupled wire construction has been discussed by Teo and Kane.\cite{teo_2014} This construction is now generalized to the Halperin bilayer states of $(m_{\uparrow}m_{\downarrow}n)$-type.

\begin{figure}
  \centering
  \includegraphics[width=\columnwidth]{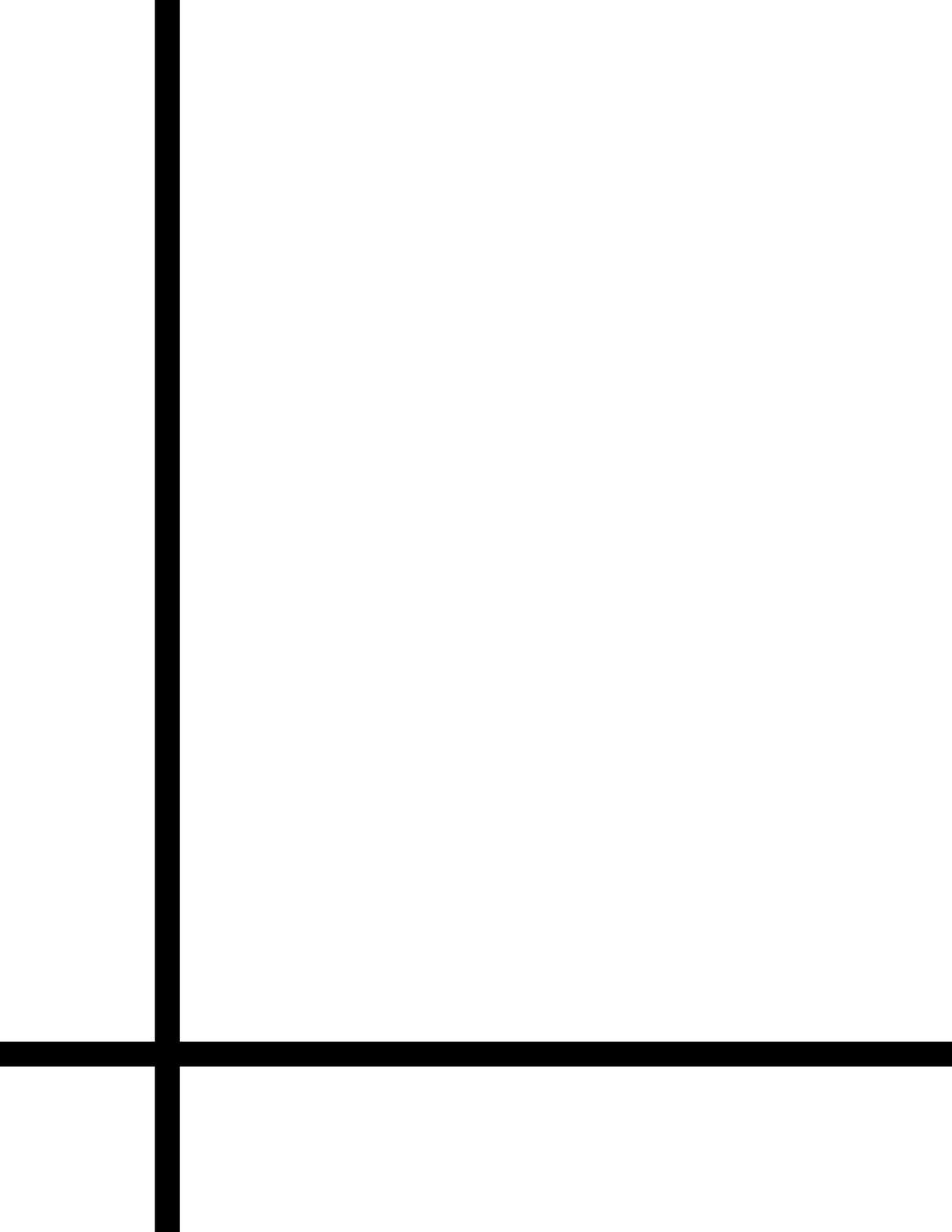}
  \caption{General form of the inter wire scatterings $g_{k+1/2\,\sigma}$ considered in the main text. The arrows in the dispersions (labeled as in Fig.~\ref{fig:dispersions}) indicate the scattering of electrons from one Fermi point to another. The integers $y$, $x_i$, $y'$, and $x_i'$ indicate how many electrons are being scattered along the corresponding arrow. Taking the Hermitian conjugate of these processes corresponds to flipping all arrows.}
  \label{fig:halperin_processes}
\end{figure}

We start from two general, local couplings involving four neighboring wires in the two layers as shown in Fig.~\ref{fig:halperin_processes}. The process denoted $g_{k+1/2\,\uparrow}$ describes a correlated tunneling of $y$ electrons with spin up between wires $k$ and $k+1$. Similarly, $g_{k+1/2\,\downarrow}$ transfers $y'$ spin down electrons between those wires.
Because of momentum conservation, not all integer values of $x_i$, $x_i'$, $y$, and $y'$ are allowed for given filling factors $\nu_\sigma$. Alternatively, a given set of $x_i$, $x_i'$, $y$, and $y'$ determines the filling factors $\nu_\sigma$ at which the corresponding processes conserve momentum. To illustrate this statement, we consider the process $g_{k+1/2\uparrow}$ written in terms of chiral fermionic fields. Using the definitions of Fig.~\ref{fig:dispersions}, we obtain

\begin{align}
g_{k+1/2\uparrow}\sim&\int dx \left(e^{ix(1-\nu_\uparrow)\Delta p}R^\dagger_{k\uparrow}(x)L^\pdag_{k+1\uparrow}(x)\right)^{y}\nonumber\\
&\times\left(e^{-ix\nu_\uparrow\Delta p}R^\dagger_{k\uparrow}(x)L^\pdag_{k\uparrow}(x)\right)^{x_1}\nonumber\\
&\times\left(e^{-ix\nu_\uparrow\Delta p}R^\dagger_{k+1\uparrow}(x)L^\pdag_{k+1\uparrow}(x)\right)^{x_2}\nonumber\\
&\times\left(e^{-ix\nu_\downarrow\Delta p}R^\dagger_{k\downarrow}(x)L^\pdag_{k\downarrow}(x)\right)^{x_3}\nonumber\\
&\times\left(e^{-ix\nu_\downarrow\Delta p}R^\dagger_{k+1\downarrow}(x)L^\pdag_{k+1\downarrow}(x)\right)^{x_4}+\text{H.c.}~.
\end{align}
This scattering is suppressed by the oscillating exponential factors unless $y (1-\nu_\uparrow) \Delta p - x_1 \nu_\uparrow \Delta p - x_2 \nu_\uparrow \Delta p - x_3 \nu_\downarrow \Delta p- x_4 \nu_\downarrow \Delta p =0$. Repeating this discussion for $g_{k+1/2\downarrow}$, we conclude that momentum conservation implies the condition

\begin{align}
\begin{pmatrix}x_1+x_2+y&x_3+x_4\\x_3'+x_4'&x_1'+x_2'+y'\end{pmatrix}
\begin{pmatrix}
\nu_{\uparrow}\\ \nu_\downarrow
\end{pmatrix} = \begin{pmatrix}
y\\y'
\end{pmatrix}~.\label{eq:halperin_momentum_cons}
\end{align}
If Eq.~\eqref{eq:halperin_momentum_cons} is satisfied, the exponential factors cancel out. The scattering $g_{k+1/2\uparrow}$, for example, then becomes
\begin{align}
g_{k+1/2\uparrow}\sim&\int dx \,R^\dagger_{k\uparrow}(x){}^{x_1+y}L^\pdag_{k\uparrow}(x){}^{x_1}\nonumber\\
&\times R^\dagger_{k+1\uparrow}(x){}^{x_2}\,L^\pdag_{k+1\uparrow}(x){}^{x_2+y}\nonumber\\
&\times\left(R^\dagger_{k\downarrow}(x)L^\pdag_{k\downarrow}(x)\right)^{x_3}\nonumber\\
&\times\left(R^\dagger_{k+1\downarrow}(x)L^\pdag_{k+1\downarrow}(x)\right)^{x_4}+\text{H.c.}~.
\end{align}
Now applying the bosonization prescription of Sec.~\ref{subsec:wire_const}, we find that the processes depicted in Fig.~\ref{fig:halperin_processes} give rise to sine-Gordon terms of the form
\begin{subequations}
\begin{align}
g_{k+1/2\,\uparrow}\sim \cos\Bigl(&(x_1+y)\Phi_{R k \uparrow}-(x_2+y)\Phi_{L k+1 \uparrow}-x_1\Phi_{L k \uparrow}\nonumber\\
&+x_2\Phi_{R k+1\uparrow}+x_3(\Phi_{Rk\downarrow}-\Phi_{Lk\downarrow})\nonumber\\
&+x_4(\Phi_{Rk+1\downarrow}-\Phi_{Lk+1\downarrow})\Bigr),   \\
g_{k+1/2\,\downarrow}\sim \cos\Bigl(&(x_1'+y')\Phi_{R k \downarrow}-(x_2'+y')\Phi_{L k+1 \downarrow}-x_1'\Phi_{L k \downarrow}\nonumber\\
&+x_2'\Phi_{R k+1\downarrow}+x_3'(\Phi_{Rk\uparrow}-\Phi_{Lk\uparrow})\nonumber\\
&+x_4'(\Phi_{Rk+1\uparrow}-\Phi_{Lk+1\uparrow})\Bigr)
\end{align}\label{eq:halp_processes}
\end{subequations}
if they preserve momentum. In the following, we search for fully gapped phases. In such a phase, all cosine perturbations $g_{k+1/2\,\sigma}$ pin the associated bosonic fields to fixed values for all $x$. This, however, requires that the arguments of the  sine-Gordon terms commute, both with themselves (at different positions) and amongst each other. For the sine-Gordon terms in Eq.~\eqref{eq:halp_processes}, we find that their arguments commute with themselves (at different positions) for

\begin{align}
x_1&=x_2~,\label{eq:cond_comm_4}\\
x_1'&=x_2'~,\label{eq:cond_comm_5}
\end{align}
and for any value of $x_3$, $x_4$, $x_3'$, $x_4'$, $y$, and $y'$. Furthermore, we find that the arguments of the couplings $g_{k+1/2\,\uparrow}$ and $g_{k+1/2\,\downarrow}$ commute if the condition

\begin{align}
y(x_3'-x_4')+y'(x_3-x_4)=0~\label{eq:cond_comm_1}
\end{align}
is satisfied (we assume $y ,y' \ne 0$). The arguments of $g_{k+1/2\,\uparrow}$ and $g_{(k\pm1)+1/2\,\downarrow}$ commute if

\begin{align}
y'x_4-yx_3'&=0~,\label{eq:cond_comm_2}\\
y'x_3-yx_4'&=0~.\label{eq:cond_comm_3}
\end{align}
Note that the combination of Eqs.~\eqref{eq:cond_comm_2} and \eqref{eq:cond_comm_3} yields Eq.~\eqref{eq:cond_comm_1}. Finally, the arguments of the couplings $g_{k+1/2\,\uparrow}$ and $g_{(k\pm1)+1/2\,\uparrow}$, and of $g_{k+1/2\,\downarrow}$ and $g_{(k\pm1)+1/2\,\downarrow}$, commute if $x_1=x_2$ and $x_1' = x_2'$, thus reproducing the condition of Eqs.~\eqref{eq:cond_comm_4} and \eqref{eq:cond_comm_5}. To study the nature of the gapless edge states associated with these couplings, it is convenient to perform the basis transformation

\begin{subequations}
\begin{align}
\widetilde{\Phi}_{Rk\uparrow} &= (x_1+y)\Phi_{R k \uparrow}-x_1\Phi_{L k \uparrow}+x_3(\Phi_{Rk\downarrow}-\Phi_{Lk\downarrow})~,\\
\widetilde{\Phi}_{Lk\uparrow} &= (x_2+y)\Phi_{L k \uparrow}-x_2\Phi_{R k\uparrow}-x_4(\Phi_{Rk\downarrow}-\Phi_{Lk\downarrow})~,\\
\widetilde{\Phi}_{Rk\downarrow} &= (x_1'+y')\Phi_{R k \downarrow}-x_1'\Phi_{L k \downarrow}+x_3'(\Phi_{Rk\uparrow}-\Phi_{Lk\uparrow})~,\\
\widetilde{\Phi}_{Lk\downarrow} &= (x_2'+y')\Phi_{L k \downarrow}-x_2'\Phi_{R k\downarrow}-x_4'(\Phi_{Rk\uparrow}-\Phi_{Lk\uparrow})~.
\end{align}\label{eq:basis_trafo}
\end{subequations}
Using the conditions of Eqs.~\eqref{eq:cond_comm_4}-\eqref{eq:cond_comm_3}, we find that these fields obey the commutation relation

\begin{align}
[\widetilde{\Phi}_{rk\sigma}(x),\widetilde{\Phi}_{r'k'\sigma'}(x')]=\delta_{rr'}\delta_{kk'}K_{\sigma\sigma'}\,i\pi r\,\text{sgn}(x'-x)~,
\end{align}
where the $K$-matrix reads

\begin{align}
K = \begin{pmatrix} y(2x_1+y)& y'(x_3+x_4)\\y'(x_3+x_4)&y'(2x_1'+y')\label{eq:kmatrix}
\end{pmatrix}~.
\end{align}
Here, we recall that $y'(x_3+x_4) = y(x_3'+x_4')$ according to Eqs.~\eqref{eq:cond_comm_2} and \eqref{eq:cond_comm_3}. In terms of these new fields, the sine-Gordon terms read
\begin{subequations}
\begin{align}
g_{k+1/2\,\uparrow}\sim \cos\Bigl(&\widetilde{\Phi}_{Rk\uparrow}-\widetilde{\Phi}_{Lk+1\uparrow}\Bigr)~,\\
g_{k+1/2\,\downarrow}\sim \cos\Bigl(&\widetilde{\Phi}_{Rk\downarrow}-\widetilde{\Phi}_{Lk+1\downarrow}\Bigr)~.
\end{align}\label{eq:cos_new_fields}
\end{subequations}
When all of these sine-Gordon terms have sufficiently large prefactors to pin their arguments to the minima of the cosines,\cite{giamarchi_book} the entire system is gapped - up to the modes $\widetilde{\Phi}_{L1\sigma}$ and $\widetilde{\Phi}_{RN\sigma}$, which simply do not have a partner field to pair up with. We have thus constructed a generalized bilayer quantum Hall state, whose gapless edge modes $\widetilde{\Phi}_{L1\sigma}$ and $\widetilde{\Phi}_{RN\sigma}$ have the $K$-matrix given in Eq.~\eqref{eq:kmatrix}. For this state to be a Halperin state of $(m_{\uparrow}m_{\downarrow}n)$-type, both the $K$-matrix given in Eq.~\eqref{eq:kmatrix}, and the matrix on the left-hand side of Eq.~\eqref{eq:halperin_momentum_cons} need to be equal to\cite{wen_book}

\begin{align}
K_{m_{\uparrow}m_\downarrow n} = \begin{pmatrix} m_\uparrow&n\\n&m_\downarrow\label{eq:kmatrixhalp}
\end{pmatrix}~.
\end{align}
This implies $y=y'=1$, $m_{\uparrow} = 2x_1+1$, $m_{\downarrow} = 2x_1'+1$ and $n=x_3+x_4$. From Eqs.~\eqref{eq:cond_comm_2} and \eqref{eq:cond_comm_3}, we furthermore find that $x_3=x_4'$ and $x_4=x_3'$ in this case.

The choice of a specific $K$-matrix, and thus of a specific Halperin state, only determines the sum of $x_3$ and $x_4$ (and of $x_3'$ and $x_4'$). The individual values of $x_3$ and $x_4$, related to the sine-Gordon terms gapping out the system, are determined by which of these sine-Gordon terms is most relevant according to renormalization group considerations.\cite{giamarchi_book} For a Hamiltonian of the form of Eq.~(\ref{eq:arrayll}), it is most favorable to subdivide $n$ as equally as possible between $x_3$ and $x_4$. For even $n$, the most relevant term has $x_3 = x_4=n/2$. For odd $n$, the system can spontaneously choose to order with either $x_3 = (n+1)/2, x_4=(n-1)/2$, or $x_3 = (n-1)/2, x_4=(n+1)/2$.

Similar to Refs.~[\onlinecite{kane_02,teo_2014}], we have assumed that the cosine perturbations are relevant operators. It is always possible\cite{kane_02} to find an appropriate Hamiltonian, incorporating local interactions between the various chiral models and generalizing Eq.~(\ref{eq:arrayll}) to reach this situation.

\subsection{Charges of a quasiparticle excitation}\label{subsec:charges}
To find additional evidence for the constructed state to be a Halperin state, we now analyze the charge of a quasiparticle excitation above the gapped bulk ground state. These excitations correspond to kinks in one of the sine-Gordon terms in the bulk, $\widetilde{\Phi}_{Rk\sigma}-\widetilde{\Phi}_{Lk+1\sigma}\to\widetilde{\Phi}_{Rk\sigma}-\widetilde{\Phi}_{Lk+1\sigma}\pm2\pi$. In order to define the charge of these excitations, we recall that the charge density of the wire $k\sigma$ is given by $\rho_{k\sigma}(x) = -\partial_x (\Phi_{R k \sigma}-\Phi_{L k \sigma})/(2\pi)$. The total charge in layer $\sigma$ is thus
\begin{align}
Q_\sigma = \frac{e}{2\pi}\sum_{k=1}^N\int dx\,\partial_x\left( \Phi_{R k \sigma}-\Phi_{L k \sigma}\right)~.\label{eq:charges_def}
\end{align}
On the other hand, Eqs.~\eqref{eq:cond_comm_4}-\eqref{eq:basis_trafo} yield

\begin{align}
\begin{pmatrix}\widetilde{\Phi}_{Rk\uparrow}-\widetilde{\Phi}_{Lk\uparrow}\\\widetilde{\Phi}_{Rk\downarrow}-\widetilde{\Phi}_{Lk\downarrow}\end{pmatrix}=\begin{pmatrix}2x_1+y&x_3+x_4\\x_3'+x_4'&2x_1'+y'\end{pmatrix}\begin{pmatrix}{\Phi}_{Rk\uparrow}-{\Phi}_{Lk\uparrow}\\{\Phi}_{Rk\downarrow}-{\Phi}_{Lk\downarrow}\end{pmatrix}~.
\end{align}
This means that

\begin{align}
\begin{pmatrix}
Q_{\uparrow}\\Q_\downarrow
\end{pmatrix} &= e \begin{pmatrix}2x_1+y&x_3+x_4\\x_3'+x_4'&2x_1'+y'\end{pmatrix}^{-1}\vec{\chi}~,\\
\vec{\chi} &= \frac{1}{2\pi} \int dx \begin{pmatrix}\partial_x\sum_k(\widetilde{\Phi}_{Rk\uparrow}-\widetilde{\Phi}_{Lk\uparrow})\\\partial_x\sum_k(\widetilde{\Phi}_{Rk\downarrow}-\widetilde{\Phi}_{Lk\downarrow})\end{pmatrix}~.
\end{align}
If we are interested in the charge associated with a kink in one of the bulk cosines, we can use $\sum_{k}(\widetilde{\Phi}_{Rk\sigma}-\widetilde{\Phi}_{Lk\sigma}) = \sum_{k \in \rm bulk}(\widetilde{\Phi}_{Rk\sigma}-\widetilde{\Phi}_{Lk+1\sigma}) + $edge terms, and therefore find that a kink in $g_{k+1/2\uparrow}$ is associated with $\vec{\chi} = (1,0)^T$, while a kink in $g_{k+1/2\downarrow}$ is associated with $\vec{\chi} = (0,1)^T$.

For the Halperin states, the two types of quasiparticle excitations have the associated charges in the two layers
\begin{align}
\begin{pmatrix}
q_{\uparrow}^{(g_{\uparrow})}\\q_{\downarrow}^{(g_{\uparrow})}
\end{pmatrix} &= e K_{m_{\uparrow}m_{\downarrow}n}^{-1}\begin{pmatrix}1\\0\end{pmatrix}~,~\begin{pmatrix}
q_{\uparrow}^{(g_{\downarrow})}\\q_{\downarrow}^{(g_{\downarrow})}
\end{pmatrix} &= e K_{m_{\uparrow}m_{\downarrow}n}^{-1}\begin{pmatrix}0\\1\end{pmatrix}~,
\end{align}
where $K_{m_{\uparrow}m_{\downarrow}n}$ is the $K$-matrix given in Eq.~\eqref{eq:kmatrixhalp}. These charges agree with the expected values for a Halperin state of $(m_{\uparrow}m_{\downarrow}n)$-type.\cite{girvin_review,degail_08}
For example, in the (331) Halperin state, the filling factors are $\nu_\uparrow = \nu_\downarrow = \frac{1}{4}$, and quasiparticles carry charges $(\frac{3e}{8}, \frac{-e}{8})$ in the two layers, giving the total charge of $e/4$.

The combination of the obtained commutation relations for the gapless edge states, the filling factors, the complete bulk gap, and the charges of quasiparticle excitations above the bulk gap finally allows one to conclude that we have indeed constructed a Halperin state via an array of coupled wires.

\section{Generalized $(m_{\uparrow}m_{\downarrow}n)$-type states in a bilayer system with opposite magnetic fields and local interactions}\label{sec:gen_halperin}
\begin{figure}
  \centering
  \includegraphics[width=\columnwidth]{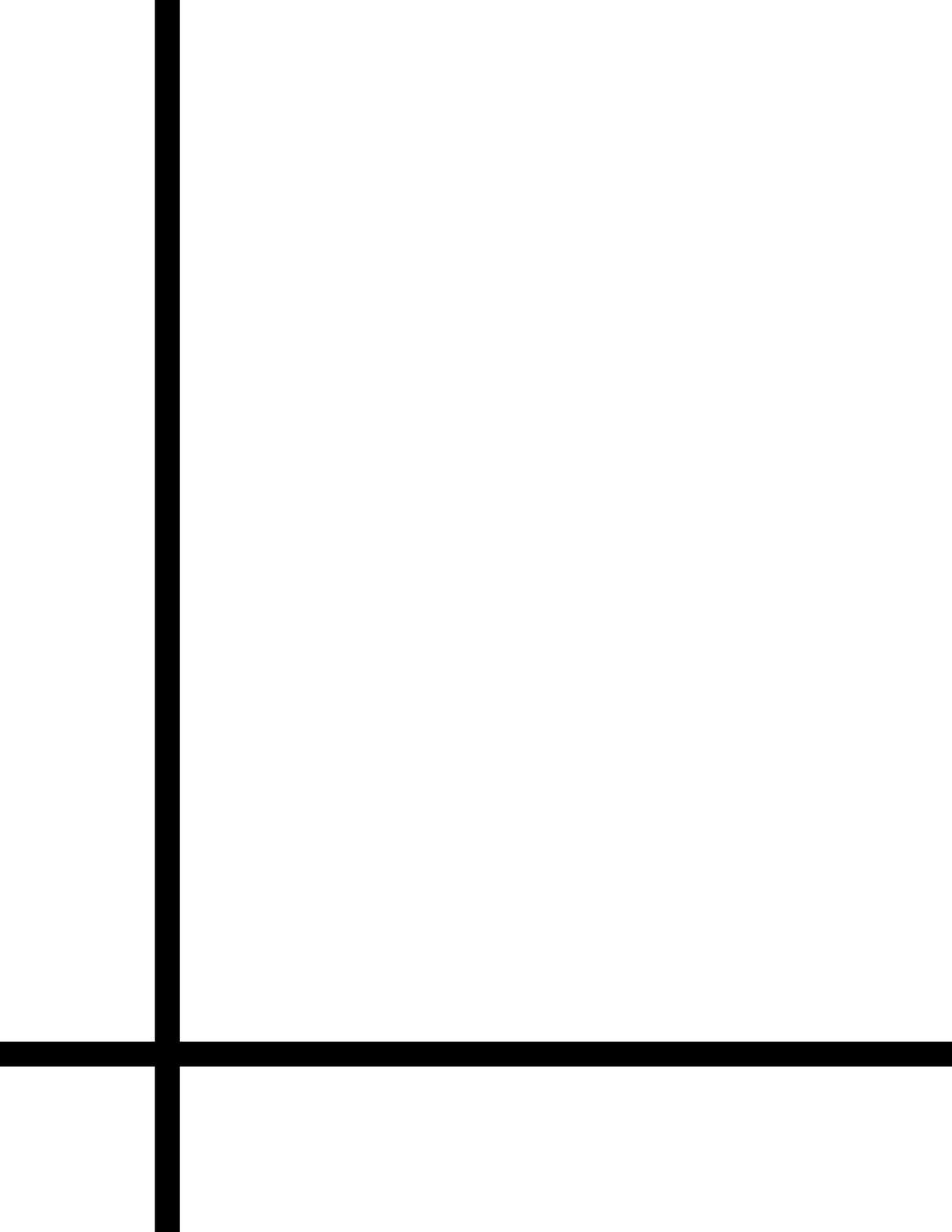}
  \caption{Dispersions for the spin-orbit-coupled case. The spin-orbit interaction results in an effective magnetic field that is opposite for the two layers (spin species). Consequently, the dispersions are shifted in opposite directions for the two spins. All labels are like in Fig.~\ref{fig:dispersions}.}
  \label{fig:dispersions_soi}
\end{figure}

In the following main part of our work, we address the coupled wire construction of topological states at zero net magnetic field for the cases where the two spin species (layers) feel an effectively opposite magnetic field. This can be realized by a spatially increasing (pseudo-) spin-orbit coupling of the form $\alpha(y) p_x\,\sigma_z$ with $\alpha(y) = \alpha_0\,y$.  For $N$ spinful wires (or a double layer with $N$ wires in each layer), this coupling gives rise to the dispersions depicted in Fig.~\ref{fig:dispersions_soi}.

\begin{figure}
  \centering
  \includegraphics[width=\columnwidth]{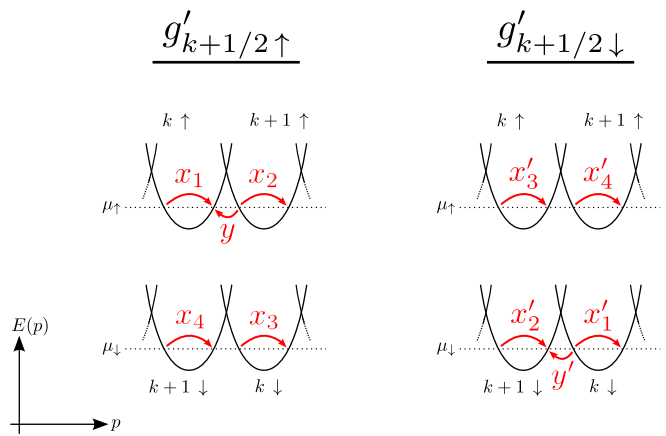}
  \caption{General form of the inter wire scatterings $g_{k+1/2\,\sigma}'$ for a spin-orbit-coupled system. All labels are as in Fig.~\ref{fig:halperin_processes}.}
  \label{fig:soi_processes}
\end{figure}

The analog of an $(m_\uparrow m_\downarrow n)$-type Halperin state in the spin-orbit-coupled system depicted in Fig.~\ref{fig:dispersions_soi} is analyzed under the important requirement of local interactions (i.e.~considering  couplings $g_{k+1/2\sigma}'$ that, for any $k$, involve only combinations of $\Phi_{rk\uparrow}$, $\Phi_{rk+1\uparrow}$, $\Phi_{rk\downarrow}$, and $\Phi_{rk+1\downarrow}$). Like in Sec.~\ref{sec:halpering_states}, we start from a general interaction process, see Fig.~\ref{fig:soi_processes}. Momentum conservation yields the same condition as given in Eq.~\eqref{eq:halperin_momentum_cons}, which we write for clarity,

\begin{align}
\begin{pmatrix}x_1+x_2+y&x_3+x_4\\x_3'+x_4'&x_1'+x_2'+y'\end{pmatrix}
\begin{pmatrix}
\nu_{\uparrow}\\ \nu_\downarrow
\end{pmatrix} = \begin{pmatrix}
y\\y'
\end{pmatrix}~.\label{eq:soi_momentum_cons}
\end{align}
If the latter equation is satisfied, bosonization yields

\begin{subequations}
\begin{align}
g_{k+1/2\,\uparrow}'\sim \cos\Bigl(&(x_1+y)\Phi_{R k \uparrow}-(x_2+y)\Phi_{L k+1 \uparrow}-x_1\Phi_{L k \uparrow}\nonumber\\
&+x_2\Phi_{R k+1\uparrow}+x_3(\Phi_{Rk\downarrow}-\Phi_{Lk\downarrow})\nonumber\\
&+x_4(\Phi_{Rk+1\downarrow}-\Phi_{Lk+1\downarrow})\Bigr)\\
g_{k+1/2\,\downarrow}'\sim \cos\Bigl(&(x_2'+y')\Phi_{R k+1 \downarrow}-(x_1'+y')\Phi_{L k \downarrow}+x_1'\Phi_{R k \downarrow}\nonumber\\
&-x_2'\Phi_{L k+1\downarrow}+x_3'(\Phi_{Rk\uparrow}-\Phi_{Lk\uparrow})\nonumber\\
&+x_4'(\Phi_{Rk+1\uparrow}-\Phi_{Lk+1\uparrow})\Bigr)~.
\end{align}\label{eq:soi_processes}
\end{subequations}
Under time reversal (TR), the wire indices transform as $k \to k$, and $\sigma \to - \sigma$. Therefore, if time reversal symmetry is satisfied, one obtains
\begin{align}
\label{eq:TR}
y= y' ,~~x_j = x'_j (j=1,2,3,4),~~~\rm{(if~TR~holds)}.
\end{align}
We will return to this condition below. The sine-Gordon terms given in Eq.~(\ref{eq:soi_processes}) differ from those in Eq.~\eqref{eq:halp_processes}. This difference stems from the fact that in order for the interaction to be local (involving only wires $k\sigma$ and $k+1\sigma$), the couplings are nonlocal in momentum space. Pictorially, drawing the process of $g_{k+1/2+M\,\uparrow}'$ in Fig.~\ref{fig:soi_processes} requires shifting the arrows $x_1,x_2,y$ to the right by $M$ wires and shifting the arrows $x_3,x_4$ to the left by $M$ wires.

Similar to the discussion of Sec.~\ref{se:couplingHalperin}, the sine-Gordon terms of Eq.~\eqref{eq:soi_processes} cannot order simultaneously for arbitrary integers $x_i$, $x_i'$, $y$, and $y'$. We find that their arguments commute with themselves at different points $x$ if

\begin{align}
x_1&=x_2~,\label{eq:cond_comm_4_soi}\\
x_1'&=x_2'~.\label{eq:cond_comm_5_soi}
\end{align}
The arguments of the couplings $g_{k+1/2\,\uparrow}'$ and $g_{k+1/2\,\downarrow}'$, on the other hand, commute if

\begin{align}
y(x_3'-x_4')+y'(x_4-x_3)=0~.\label{eq:cond_comm_1_soi}
\end{align}
In addition, we find that the arguments of $g_{k+1/2\,\uparrow}$ and $g_{(k\pm1)+1/2\,\downarrow}$ commute if

\begin{align}
y'x_4+yx_3'&=0~,\label{eq:cond_comm_2_soi}\\
y'x_3+yx_4'&=0~.\label{eq:cond_comm_3_soi}
\end{align}
Again, the combination of Eq.~\eqref{eq:cond_comm_2_soi} and Eq.~\eqref{eq:cond_comm_3_soi} yields Eq.~\eqref{eq:cond_comm_1_soi}. Finally, the arguments of $g_{k+1/2\,\uparrow}'$ and $g_{(k\pm1)+1/2\,\uparrow}'$, and of $g_{k+1/2\,\downarrow}'$ and $g_{(k\pm1)+1/2\,\downarrow}'$, commute if $x_1=x_2$, and $x_1'=x_2'$. Note that we have again assumed $y ,y' \ne 0$. Importantly, we thus find that not only the sine-Gordon terms, but also the conditions following from the commutation relations are different from those obtained in Sec.~\ref{sec:halpering_states}.

To study the edge modes of this state, it is again helpful to define new fields

\begin{subequations}
\begin{align}
\widetilde{\Phi}_{Rk\uparrow}' &= (x_1+y)\Phi_{R k \uparrow}-x_1\Phi_{L k \uparrow}+x_3(\Phi_{Rk\downarrow}-\Phi_{Lk\downarrow})~,\\
\widetilde{\Phi}_{Lk\uparrow}' &= (x_2+y)\Phi_{L k \uparrow}-x_2\Phi_{R k\uparrow}-x_4(\Phi_{Rk\downarrow}-\Phi_{Lk\downarrow})~,\\
\widetilde{\Phi}_{Rk\downarrow}' &= (x_1'+y')\Phi_{L k \downarrow}-x_1'\Phi_{R k \downarrow}-x_3'(\Phi_{Rk\uparrow}-\Phi_{Lk\uparrow})~,\\
\widetilde{\Phi}_{Lk\downarrow}' &= (x_2'+y')\Phi_{R k \downarrow}-x_2'\Phi_{L k\downarrow}+x_4'(\Phi_{Rk\uparrow}-\Phi_{Lk\uparrow})~,
\end{align}\label{eq:basis_trafo_soi}
\end{subequations}
which satisfy
\begin{align}
\label{commu}
[\widetilde{\Phi}_{rk\sigma}'(x),\widetilde{\Phi}_{r'k'\sigma'}'(x')]=\delta_{rr'}\delta_{kk'}K_{\sigma\sigma'}'\,i\pi r\,\text{sgn}(x'-x)~,
\end{align}
where the $K$-matrix reads

\begin{align}
K' = \begin{pmatrix} y(2x_1+y)& -y(x_3'+x_4')\\y'(x_3+x_4)&-y'(2x_1'+y')\label{eq:kmatrix_soi}
\end{pmatrix}~,
\end{align}
with $y'(x_3+x_4) = -y(x_3'+x_4')$ according to Eqs.~\eqref{eq:cond_comm_2_soi} and \eqref{eq:cond_comm_3_soi}. Expressed in these new fields, the sine-Gordon terms read
\begin{subequations}
\begin{align}
g_{k+1/2\,\uparrow}'\sim \cos\Bigl(&\widetilde{\Phi}_{Rk\uparrow}'-\widetilde{\Phi}_{Lk+1\uparrow}'\Bigr)~,\\
g_{k+1/2\,\downarrow}'\sim \cos\Bigl(&\widetilde{\Phi}_{Rk\downarrow}'-\widetilde{\Phi}_{Lk+1\downarrow}'\Bigr)~.
\end{align}\label{eq:cos_new_fields_soi}
\end{subequations}
The bilayer thus has the fields $\widetilde{\Phi}_{L1\sigma}'$ and $\widetilde{\Phi}_{RN\sigma}'$ as gapless edge modes characterized by the non-trivial $K$-matrix given in Eq.~\eqref{eq:kmatrix_soi}.

The commutation relation given in Eq.~\eqref{commu} indicates that the fields $\widetilde{\Phi}_{Rk\downarrow}'$ actually represent left moving modes, while $\widetilde{\Phi}_{Lk\downarrow}'$ correspond to right movers. At the left edge (near the $k=1$ wire), this implies the existence of one gapless left moving mode $\widetilde{\Phi}_{L1\uparrow}'$, and one gapless right moving mode $\widetilde{\Phi}_{L1\downarrow}'$. Similarly, the left and right moving modes propagating in the right edge, namely near the $k=N$ wire, are $\widetilde{\Phi}_{RN\downarrow}'$ and $\widetilde{\Phi}_{RN\uparrow}'$, respectively. Thus, the $L/R$ indices in Eq.~(\ref{eq:basis_trafo_soi}) actually do not mark the chirality of edge states, but rather the edge at which they live.

To define the analogue of an $(m_\uparrow m_\downarrow n)$ state, we set $y=y'=1$. With this choice, Eqs.~\eqref{eq:cond_comm_2_soi} and \eqref{eq:cond_comm_3_soi} yield $x_3=-x_4'$ and $x_4=-x_3'$. Using $m_{\uparrow} = 2x_1+1$, $m_{\downarrow} = 2x_1'+1$ and $n=x_3+x_4$, we obtain the $K$-matrix
\begin{align}
K_{m_\uparrow m_\downarrow n}' = \begin{pmatrix} m_\uparrow& n\\n&-m_\downarrow\label{eq:kmatrix_soi_2rep}~
\end{pmatrix}.
\end{align}
One may observe that in the presence of TR, when Eq.~(\ref{eq:TR}) holds, the solutions to Eqs.~(\ref{eq:cond_comm_2_soi}) and (\ref{eq:cond_comm_3_soi}) yield $x_3 = -x_4$. TR states thus necessarily have $n=0$.

We finally note that similar to the discussion in Sec.~\ref{se:couplingHalperin}, we can always find a Luttinger liquid Hamiltonian such that the cosine perturbations are relevant.

\subsection{Charges of quasiparticle excitations}
As in  Sec.~\ref{subsec:charges}, a quasiparticle excitation above the bulk gap corresponds to a kink in one of the bulk cosines, $\widetilde{\Phi}_{Rk\sigma}'-\widetilde{\Phi}_{Lk+1\sigma}'\to\widetilde{\Phi}_{Rk\sigma}'-\widetilde{\Phi}_{Lk+1\sigma}'\pm 2\pi$. Their charges are defined by Eq.~\eqref{eq:charges_def}. Using Eqs.~\eqref{eq:cond_comm_4_soi} - \eqref{eq:basis_trafo_soi}, we obtain

\begin{align}
\label{eq:rho}
\begin{pmatrix}\widetilde{\Phi}_{Rk\uparrow}'-\widetilde{\Phi}_{Lk\uparrow}'\\\widetilde{\Phi}_{Rk\downarrow}'-\widetilde{\Phi}_{Lk\downarrow}'\end{pmatrix}=
\hat{M}\begin{pmatrix}{\Phi}_{Rk\uparrow}-{\Phi}_{Lk\uparrow}\\{\Phi}_{Rk\downarrow}-{\Phi}_{Lk\downarrow}\end{pmatrix},
\end{align}
where
\begin{align}
\hat{M}=\begin{pmatrix}2x_1+y&x_3+x_4\\-(x_3'+x_4')&-(2x_1'+y')\end{pmatrix}.
\end{align}
This implies that the charge components in the two layers of bulk quasiparticles associated with a kink in $g_{k+1/2\uparrow}$ are given by

\begin{align}
\begin{pmatrix}
q_{\uparrow}^{(g_{\uparrow}')}\\q_{\downarrow}^{(g_{\uparrow}')}
\end{pmatrix} &= \hat{M}^{-1}\begin{pmatrix}1\\0\end{pmatrix}~,
\end{align}
while an (anti-)kink in $g_{k+1/2\downarrow}$ is associated with charges

\begin{align}
\begin{pmatrix}
q_{\uparrow}^{(g_{\downarrow}')}\\q_{\downarrow}^{(g_{\downarrow'})}
\end{pmatrix} &= e \hat{M}^{-1}\begin{pmatrix}0\\-1\end{pmatrix}~.
\end{align}
As an example, the state generated by $y=y'=1$, $x_3=-x_4'$, $x_4=-x_3'$ has quasiparticles with charges

\begin{align}
\begin{pmatrix}
q_{\uparrow}^{(g_{\uparrow}')}\\q_{\downarrow}^{(g_{\uparrow}')}
\end{pmatrix} &= \frac{e}{m_\uparrow m_\downarrow+n^2}\begin{pmatrix}m_\downarrow\\n\end{pmatrix}~,\label{eq:qgu}
\end{align}
and
\begin{align}
\begin{pmatrix}
q_{\uparrow}^{(g_{\downarrow}')}\\q_{\downarrow}^{(g_{\downarrow}')}
\end{pmatrix} &= \frac{e}{m_\uparrow m_\downarrow+n^2}\begin{pmatrix}-n\\m_\uparrow\end{pmatrix}~.\label{eq:qgd}
\end{align}
Note that for $n \ne 0$, the two types of quasiparticles in general have different total charge. For example, in the (331) state one quasiparticle has charge $ \frac{2}{5}e$, and the other one has charge $\frac{1}{5}e$.

\subsection{Quantum Hall conductivity}
To compute the Hall conductivity, we put the system in a Corbino geometry. Following the Laughlin argument, the adiabatic insertion of a flux quantum leads to a charge $Q$ being pumped between the inner and outer edges, which is related to the Hall conductivity by $\sigma_{xy} = \frac{e}{h}Q$. This can be directly computed from the $K$-matrix,\cite{wen_book} but we give the derivation here for completeness.

We write the Hamiltonian of the $k=1$ edge including the coupling to the electromagnetic field using the expressions in Eqs.~(\ref{eq:charges_def}) and (\ref{eq:rho}) for the density,
\begin{equation}
H = \partial_x{\vec{\phi}} \hat{V}  \partial_x{\vec{\phi}} + \frac{1}{2 \pi}   [(1,1) \cdot  \hat{M}^{-1}    \vec{\phi} ]\epsilon_{\mu \nu} \partial_\mu A_\nu,
\end{equation}
with $\vec{\phi}_i = (\widetilde{\Phi}_{L1\uparrow}',\widetilde{\Phi}_{L1\downarrow}')$, and where $\hat{V}$ is a generic term containing information on the velocity of the two counter propagating modes as well as the interaction between them. We now consider a time dependent flux $\Phi(t)$ inserted through the hole in the Corbino geometry, giving rise to an electric field $\epsilon^{\mu \nu} \partial_\mu A_\nu = E = \frac{\partial_t \Phi}{L}$, where $L$ is the circumference.

We can write the Heisenberg equations of motion using the commutation relations of Eq.~(\ref{commu}),
\begin{equation}
\partial_t \partial_x \vec{\phi}  = \frac{\partial_t \Phi}{L} [  K'^T ({\hat{M}}^{-1})^T \cdot (1,1)^T] - 4 \pi [K'^T V^T \partial^2_x \vec{\phi}].
\end{equation}
An integration over space, and the use of Eqs.~(\ref{eq:charges_def}) and (\ref{eq:rho}) yields the charge pumped into the $k=1$ edge, which must come from the $K=N$ edge,
\begin{equation}
\partial_t Q   = e [(1,1) \cdot \hat{M}^{-1} {K'}^T ({\hat{M}}^{-1})^T(1,1)^T] \frac{\partial_t \Phi}{2 \pi}.
\end{equation}
Notice that the $\hat{V}$ term gives a full derivative and can be neglected assuming that $\partial_x \vec{\phi}$ is a constant in the ground state. Concentrating on an $(m_\uparrow m_\downarrow n)$ state, we find that the adiabatic insertion of a $2 \pi$ flux results in a total pumped charge of $\frac{2 n +m_\downarrow - m_\uparrow}{m_\uparrow m_\downarrow + n^2}$, corresponding to the Hall conductivity
\begin{equation}
\sigma_{xy}= \frac{e^2}{h}  \frac{2n + m_\downarrow-m_\uparrow}{m_\uparrow m_\downarrow + n^2}.
\end{equation}
As required, the Hall response vanishes in the time reversal symmetric case, $m_\uparrow = m_\downarrow$, $n=0$. It is, however, generically finite for the class of $(m_\uparrow m_\downarrow n)$ states under consideration. Specifically, $\sigma_{xy}$ depends on the sign of $n$. In the special case $m_\uparrow = m_\downarrow$, the sign of $\sigma_{xy}$ is determined by that of $n$. We thus find that like a finite spin polarization, a finite Hall conductivity is a clear indicator of a TR broken topological insulating phase.

For a system with conserved $s_z$, one may discuss the Hall conductivity of each spin separately. We find that the pumped charges are
\begin{align}
\partial_t Q_\uparrow   = e [(1,0) \cdot \hat{M}^{-1} {K'}^T ({\hat{M}}^{-1})^T(1,1)^T] \frac{\dot{\Phi}}{2 \pi},  \nonumber\\
\partial_t Q_\downarrow   = e [(0,1) \cdot \hat{M}^{-1} {K'}^T ({\hat{M}}^{-1})^T(1,1)^T] \frac{\dot{\Phi}}{2 \pi},\label{eq:charges_eom}
\end{align}
From this, the spin Hall conductivity is found to be
\begin{align}
\sigma_{sH} =&\frac{e}{4 \pi} [(1,-1) \cdot \hat{M}^{-1} {K'}^T ({\hat{M}}^{-1})^T(1,1)^T]  \nonumber\\
=& \frac{e}{4 \pi} \frac{m_\uparrow + m_\downarrow   }{m_\uparrow m_\downarrow + n^2}.
\end{align}
We note that the charges $Q_\sigma$ obtained via the equations of motion technique, see Eqs.~\eqref{eq:charges_eom}, equal the total charges for a combined kink in both types of bulk cosines, which are given in Eqs.~\eqref{eq:qgu} and \eqref{eq:qgd}. The quantum Hall and spin Hall conductivities can thus be understood as describing the transport of both types of quasiparticles across the sample.

\subsection{Stability of the edge}
The present situation with counter propagating edge modes raises the question on their stability against impurity scattering. In the presence of an impurity, for instance near the left edge, one may write a term of the form
\begin{equation}
H_{\rm imp} \sim g_{\rm imp } \cos( \widetilde{\Phi}_{L1\uparrow}' - \widetilde{\Phi}_{L1\downarrow}') \label{eq:gap_edge_proc}
\end{equation}
This type of perturbation does not conserve spin. We do not, however, expect to have such a conservation in a generic system with spin orbit coupling, and even less so in the presence of broken time reversal symmetry. Our specific model has an additional issue: The momentum difference between the states described by $\widetilde{\Phi}_{L1\uparrow}'$ and $\widetilde{\Phi}_{L1\downarrow}'$ grows with the width of the system. We believe that this is just a property of the specific model. In Sec.~\ref{sec:negmass}, we consider an alternative model in which the issue of large momentum difference between edge modes is absent. Similarly, this issue is absent in the model presented in Ref.~[\onlinecite{sagi_14}]. We hence conclude that the edge modes are generically unstable. Yet, the bulk gap and its underlying topological properties, including Hall response and quasiparticles, are unaffected by this edge physics.

\subsection{Negative masses in second layer}
\label{sec:negmass}
\begin{figure}
  \centering
  \includegraphics[width=\columnwidth]{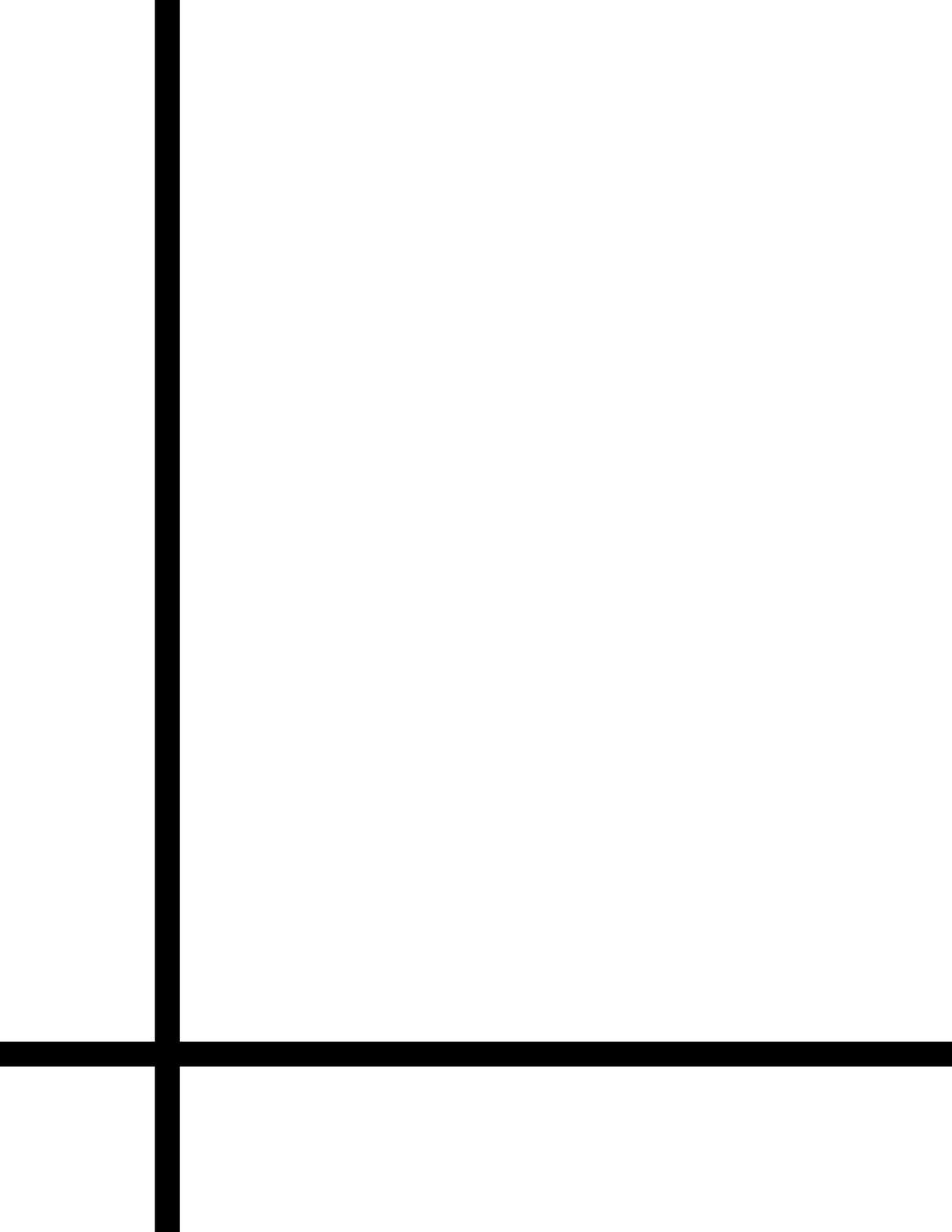}
  \caption{Dispersions for the a double layer with negative mass in the lower layer, and in the presence of a magnetic field. The filling $\nu_\uparrow$ corresponds to the filling of electrons in the upper layer, while $\nu_\downarrow$ is the filling of holes in the lower layer. All other labels are as in Fig.~\ref{fig:dispersions}.}
  \label{fig:dispersions_neg_mass}
\end{figure}

\begin{figure}
  \centering
  \includegraphics[width=\columnwidth]{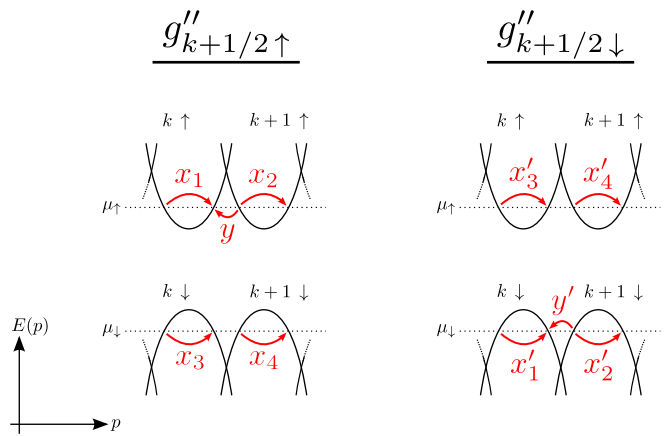}
  \caption{General form of the inter wire scatterings $g_{k+1/2\,\sigma}''$ for a double layer system with opposite masses. All labels are as in Fig.~\ref{fig:halperin_processes}.}
  \label{fig:neg_mass_processes}
\end{figure}

Instead of using two layers of quantum wires with an effectively opposite magnetic field, local interactions can also stabilize a (fractional) topological insulator state in a double layer of wires subject to a homogenous magnetic field if the masses in the two layers are opposite.\cite{sagi_14,klinovaja_14} This gives rise to the dispersions depicted in Fig.~\ref{fig:dispersions_neg_mass}. When analyzing the general interaction processes $g_{k+1/2\sigma}''$ shown in Fig.~\ref{fig:neg_mass_processes}, we obtain the same conditions as for the spin-orbit-coupled system with identical masses, namely Eqs.~\eqref{eq:soi_momentum_cons} and \eqref{eq:cond_comm_4_soi}-\eqref{eq:cond_comm_3_soi}. This implies that also for layers with opposite masses, a Halperin-like state with sufficiently strong interlayer correlations can form a time reversal broken state. As advertised in the last subsection, a process of the form of Eq.~\eqref{eq:gap_edge_proc} now preserves momentum, and is thus susceptible to gap the edge states. To see this, we remark that for negative masses in the second layer, the modes $R_{1\uparrow}$, $L_{1\uparrow}$ live at the same momenta than the modes $R_{1\downarrow}$, $L_{1\downarrow}$ (see Fig.~\ref{fig:dispersions_neg_mass}). This was not the case in the setup discussed before, whose dispersions are shown in Fig.~\ref{fig:dispersions_soi}.

\section{Conclusions and outlook}
\label{sec:conclusion}
In this work, we have identified and analyzed a generalized class of quantum Hall states at zero magnetic field, which can be understood as analogues of fractional topological insulator states with strong correlations between time-reversal partners. Importantly, we find that these states have broken time reversal symmetry, which in general can lead to spin magnetization or to a non-zero Hall conductance, or  to both. This suggests the possibility that fractional topological insulators may be unstable towards the formation of a time-reversal symmetry broken state in the presence of sufficiently strong electron-electron interactions.

To close our discussion, let us briefly comment on the consistency of our findings with other approaches for fractional topological insulators. Refs.~[\onlinecite{levin_stern_09,neupert_11,santos_11,levin_stern_12}] have analyzed Chern-Simons theories of time reversal symmetric fractional topological insulators, and inferred that the $K$-matrices of these theories should satisfy a number of properties. For a $(2\times 2)$ $K$-matrix, the only possibility is to have vanishing off-diagonal elements, which corresponds to two decoupled copies of a fractional quantum Hall state. A time reversal symmetric state with off-diagonal elements in the $K$-matrix, encoding correlations between the time reversal partners, requires the $K$-matrix to be larger than $(2\times 2)$. 

This is consistent with the family of states that we have found. These states are characterized by the $(2\times 2)$ $K$-matrix given in Eq.~(\ref{eq:kmatrix_soi_2rep}), in which non-vanishing off-diagonal elements imply broken time reversal symmetry. Being characterized by a relatively simple $K$-matrix, and not a larger one as discussed for time reversal symmetric states with inter spin correlations,\cite{levin_stern_09,neupert_11,santos_11,levin_stern_12} the states analyzed in this work form alternative candidate time reversal symmetry broken states that should be considered in the on-going search for fractional topological insulators in different models.

One may also ask how our findings fit into the general classification scheme of Ref.~[\onlinecite{neupert_14}]. While the states with time reversal symmetry discussed here (the $n=0$ or $(m,m,0)$ states) are simply to those analyzed in Ref.~[\onlinecite{neupert_14}] within symmetry class $AII$, our spin up-spin down interacting states with $n \ne 0$ belong to symmetry class $A$. However, they were not discussed in Ref.~[\onlinecite{neupert_14}], which was in this class restricted to the simplest states with single component $(1\times 1)$ $K-$matrices.

The model presented here is strongly anisotropic. Yet, we believe that the resulting phases are ground states of isotropic systems, similar to the wire construction of the fractional quantum Hall effect which gives the same physical state described by the Laughlin circular symmetric wave function.

Few interesting questions and directions remain to be explored. Whereas here we have mainly elaborated on the classification of a family of TR broken states, a microscopic study of a specific model, as well as proposed realizations which can show the emergence of such symmetry broken phases, have not been included in this paper and are left for a future study. Also the theory of the phase transition is an interesting issue which was not analyzed here. Indeed one may consider the spin-polarization, or the Hall conductivity, both of which vanish in the symmetric phase, and construct an appropriate order parameter. However, a conventional Ginzburg Landau theory for spontaneous symmetry breaking can not be sufficient by itself, since an order parameter expressed in terms of the spin-polarization, or the Hall conductivity, becomes quantized in the ordered phase. Thus, symmetry breaking and topology conspire to yield a fractionally quantized order parameter. Understanding the nature of the phase transition is left for a future study.

 \acknowledgements
We thank Lars Fritz, Moshe Goldstein, Yuval Oreg, Eran Sagi and Ady Stern for useful discussions.  This work has been supported by Swiss NF, and NCCR QSIT (TM), and ISF and Marie Curie CIG grants (ES). TM gratefully acknowledges the hospitality of the Tel Aviv University, where parts of this work have been performed.

%%%%%%%%%%%%%%%%%%%%%%%


\begin{thebibliography}{99}

\bibitem{kane_mele_05}
C. L. Kane and E. J. Mele, Phys. Rev. Lett. \textbf{95}, 226801 (2005).

\bibitem{bernevig_zhang_2006}
B. A. Bernevig and S.-C. Zhang, Phys. Rev. Lett. \textbf{96}, 106802 (2006).

\bibitem{bernevig_hughes_zhang_2006}
B. A. Bernevig, T. L. Hughes and S.-C. Zhang, Science \textbf{314}, 1757 (2006).

\bibitem{liu_08}
C. Liu, T. L. Hughes, X.-L. Qi, K. Wang, and S.-C. Zhang, Phys. Rev. Lett. \textbf{100}, 236601 (2008).

\bibitem{ti_review}
M. Z. Hasan and C. L. Kane, Rev. Mod. Phys. \textbf{82}, 3045 (2010).

\bibitem{koenig_07}
M. K\"{o}nig, S. Wiedmann, C. Br\"{u}ne, A. Roth, H. Buhmann, L. W. Molenkamp, X.-L. Qi, and S.-C. Zhang, Science \textbf{318}, 766 (2007).

\bibitem{knez_11}
I. Knez, R.-R. Du, and G. Sullivan, Phys. Rev. Lett. \textbf{107}, 136603 (2011).

\bibitem{du_13}
L. Du, I. Knez, G. Sullivan, and R.-R. Du, arXiv:1306.1925.

\bibitem{levin_stern_09}
M. Levin and A. Stern, Phys. Rev. Lett. \textbf{103}, 196803 (2009).

\bibitem{Freedman}
M. Freedman, C. Nayak, K. Shtengel, K. Walker, and Z. Wang, Ann. Phys. \textbf{310}, 428 (2004).

\bibitem{neupert_11}
T. Neupert, L. Santos, S. Ryu, C. Chamon, and C. Mudry, Phys. Rev. B \textbf{84}, 165107 (2011).

\bibitem{santos_11}
L. Santos, T. Neupert, S. Ryu, C. Chamon, and C. Mudry, Phys. Rev. B \textbf{84}, 165138 (2011).

\bibitem{levin_stern_12}
M. Levin and A. Stern, Phys. Rev. B \textbf{86}, 115131 (2012).

\bibitem{poilblanc_87}
D. Poilblanc, G. Montambaux, M. H\'{e}ritier, and P. Lederer, Phys. Rev. Lett. \textbf{58}, 270 (1987).

\bibitem{yakovenko_91}
V. M. Yakovenko, Phys. Rev. B \textbf{43}, 11353 (1991).

\bibitem{sondhi_00}
S. L. Sondhi and K. Yang, Phys. Rev. B \textbf{63}, 054430 (2001).

\bibitem{kane_02}
C. L. Kane, R. Mukhopadhyay, and T. C. Lubensky, Phys. Rev. Lett. \textbf{88}, 036401 (2002).

\bibitem{teo_2014}
J. C. Y. Teo and C. L. Kane, Phys. Rev. B \textbf{89}, 085101 (2014).

\bibitem{GANGOF11}
R. S. K. Mong et al., Phys. Rev. X \textbf{4}, 011036 (2014).

\bibitem{OregSelaStern}
Y. Oreg, E. Sela, A. Stern, Phys. Rev. B \textbf{89}, 115402 (2014).

\bibitem{klinovaja_epjb_14}
J. Klinovaja and D. Loss, Eur. Phys. J. B \textbf{87}, 171 (2014).

\bibitem{sagi_14}
E. Sagi and Y. Oreg, Phys. Rev. B \textbf{90}, 201102(R) (2014).

\bibitem{klinovaja_14}
J. Klinovaja and Y. Tserkovnyak, Phys. Rev. B \textbf{90}, 115426 (2014).

\bibitem{neupert_14}
T. Neupert, C. Chamon, C. Mudry, and R. Thomale, Phys. Rev. B \textbf{90}, 205101 (2014).


\bibitem{auslaender_02}
O. M. Auslaender, A. Yacoby, R. de Picciotto, K. W. Baldwin, L. N. Pfeiffer, and K. W. West, Science \textbf{295}, 825 (2002).

\bibitem{wen_book}
C. G. Wen, {\it Quantum Field Theory of Many-Body Systems} (Oxford University Press, New York, 2007).

\bibitem{halperin_83}
B. I. Halperin, Helv. Phys. Acta \textbf{56}, 75 (1983).

\bibitem{haldane_83}
F. D. M. Haldane, Phys. Rev. Lett. \textbf{51}, 605 (1983).

\bibitem{halperin_84}
B. I. Halperin, Phys. Rev. Lett. \textbf{52}, 1583 (1984).

\bibitem{degail_08}
R. de Gail, N. Regnault, and M. O. Goerbig, Phys. Rev. B \textbf{77}, 165310 (2008).

\bibitem{giamarchi_book}
T. Giamarchi, {\it Quantum Physics in One Dimension} (Oxford University Press, New York, 2003).

\bibitem{girvin_review}
S. M. Girvin and A. H. MacDonald, in {\it Perspectives in Quantum Hall Effects}, edited by S. Das Sarma and A. Pinczuk (Wiley, New York, 1997).






\end{thebibliography}
\end{document}